\documentclass[apj]{emulateapj}
\usepackage{pstricks,amsmath}
\usepackage{apjfonts}

\def\1e{\mbox{1E\,0657--56}}
\def\kms        {km$\;$s$^{-1}$}
\def\hseventy   {$H_0=70$~km$\;$s$^{-1}\,$Mpc$^{-1}$}
\def\chandra    {\emph{Chandra}}
\def\vla        {\emph{VLA}}
\def\as         {$^{\prime\prime}$}
\def\lax{\lesssim}

\def\bi{\bfseries\itshape}

\begin{document}

\submitted{ApJ in press, submitted 2004 December 16, accepted 2005 March 30}

\lefthead{BOW SHOCK IN A520}
\righthead{MARKEVITCH ET AL.}

\title{BOW SHOCK AND RADIO HALO IN THE MERGING CLUSTER A520}

\author{M.~Markevitch\altaffilmark{1,2}, 
F.~Govoni\altaffilmark{3},
G.~Brunetti\altaffilmark{3}, and
D.~Jerius\altaffilmark{1}}

\altaffiltext{1}{Harvard-Smithsonian Center for Astrophysics, 60 Garden St.,
Cambridge, MA 02138; maxim@head.cfa.harvard.edu}

\altaffiltext{2}{Space Research Institute, Moscow, Russia}

\altaffiltext{3}{Istituto di Radioastronomia del CNR, Bologna, Italy}

\setcounter{footnote}{3}

\begin{abstract}

\chandra\ observations of the merging galaxy cluster A520 reveal a prominent
bow shock with $M=2.1^{+0.4}_{-0.3}$. This is only the second clear example
of a substantially supersonic merger shock front in clusters.  Comparison of
the X-ray image with that of the previously known radio halo reveals a
coincidence of the leading edge of the halo with the bow shock, offering an
interesting experimental setup for determining the role of shocks in the
radio halo generation. The halo in A520 apparently consists of two spatially
distinct parts, the main turbulence-driven component and a cap-like forward
structure related to the shock, where the latter may provide pre-energized
electrons for subsequent turbulent re-acceleration.  The radio edge may be
caused by electron acceleration by the shock. If so, the synchrotron
spectrum should have a slope of $\alpha\simeq 1.2$ right behind the edge
with quick steepening further away from the edge.  Alternatively, if shocks
are inefficient accelerators, the radio edge may be explained by an increase
in the magnetic field and density of pre-existing relativistic electrons due
to gas compression.  In the latter model, there should be radio emission in
front of the shock with the same spectrum as that behind it, but 10--20
times fainter.  If future sensitive radio measurements do not find such
pre-shock emission, then the electrons are indeed accelerated (or
re-accelerated) by the shock, and one will be able to determine its
acceleration efficiency. We also propose a method to estimate the magnetic
field strength behind the shock, based on measuring the dependence of the
radio spectral slope upon the distance from the shock. In addition, the
radio edge provides a way to constrain the diffusion speed of the
relativistic electrons.

\end{abstract}


\section{INTRODUCTION}
\label{sec:intro}

Cluster mergers convert kinetic energy of the gas in colliding subclusters
into thermal energy by driving shocks and turbulence in the gaseous halo of
the merged cluster. A fraction of this energy may be diverted into
nonthermal phenomena, such as magnetic field amplification and the
acceleration of relativistic particles that manifest themselves via
synchrotron radio halos (recently reviewed by, e.g., Feretti 2002, 2004;
Kempner et al.\ 2004) and inverse Compton hard X-ray emission (e.g.,
Fusco-Femiano et al.\ 2004; Rephaeli \& Gruber 2002).  This energy fraction
depends on microphysics of the magnetized intracluster plasma, which is
poorly known. Shock fronts may provide a unique observational tool to study
the above processes, because they create high-contrast features in the
cluster X-ray images (and, as we shall see, in the radio images), and
because they are those rare locations in clusters where gas velocities in
the sky plane can be determined from the X-ray imaging spectroscopy (e.g.,
Markevitch, Sarazin, \& Vikhlinin 1999).

Until recently, the only unambiguous cluster merger shock (exhibiting both a
sharp gas density edge and a convincing temperature jump, the prerequisites
for determining the gas velocity and density jumps) was that found by
\chandra\ in \1e\ (Markevitch et al.\ 2002). In this paper, we present the
second such example discovered in A520, a merging cluster at $z=0.203$.
While many clusters exhibit merging and shock-heated gas, observations of
shock \emph{fronts} are so rare because one has to catch a merger at a very
specific stage when the shock has not yet moved to the outer, low surface
brightness regions, and at a sufficiently small angle from the sky plane so
that projection does not hide the density edge.

Both A520 and \1e\ have radio halos (Govoni et al.\ 2001; Liang et al.\
2000).  Close relation between halos and mergers was extensively discussed
(e.g., Feretti 2002; Buote 2001; Markevitch \& Vikhlinin 2001; Govoni et
al.\ 2004, hereafter G04).  Merger-driven turbulence is likely the main
process responsible for generation of the ultrarelativistic electrons
producing the diffuse radio emission (e.g., Schlickeiser, Sievers, \&
Thiemann 1987; Brunetti et al.\ 2001; Fujita, Takizawa, \& Sarazin 2003),
but the contribution due to shock acceleration (e.g., Harris et al.\ 1980;
Tribble 1993; Sarazin 1999) is still not clear. The best way to separate
these two contributions is to look at a cluster with a bow shock and a radio
halo, which is what we do in this paper.

We assume a flat cosmology with \hseventy\ and $\Omega_0=0.3$, in which 1\as\
is 3.34 kpc at the cluster's redshift. Uncertainties are 90\% unless stated
otherwise.

%%%%%%%%%%%%%%%%%%%%%%%%%%%%%%%%%%%%%%%%%%%%%%%%%%%%%%%%%%%%%%%%%%%%%%%%%%
\begin{figure*}[t]
\pspicture(0,15.5)(18.5,24.0)
%\psgrid(0,15)(18.,24)

\rput[tl]{0}(0.8,24){\epsfysize=7.9cm \epsfclipon
\epsffile[3 5 536 533]{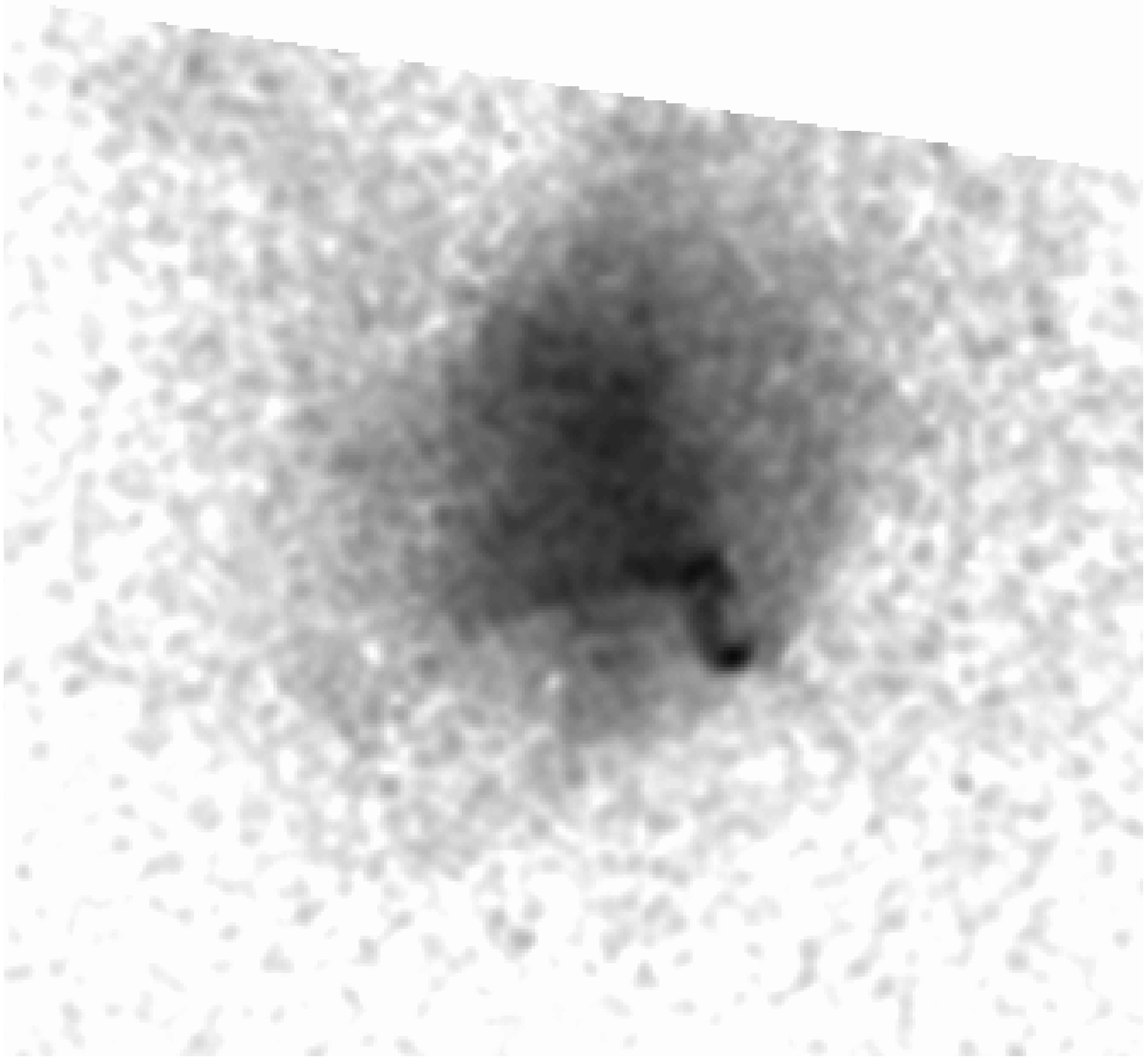}}

\rput[tl]{0}(9.4,24){\epsfysize=7.9cm \epsfclipon
\epsffile[3 5 536 533]{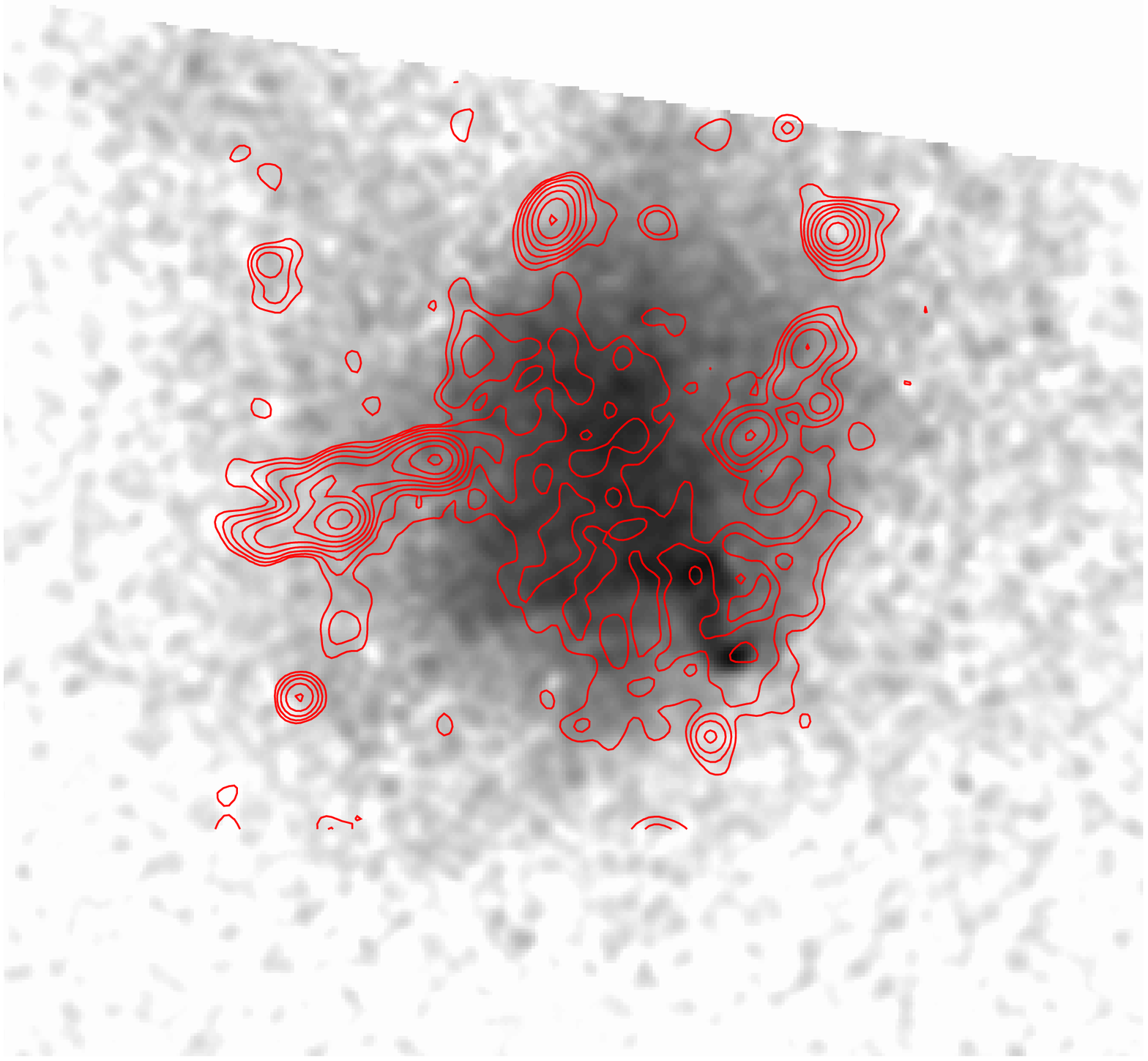}}

\rput[tl]{0}(0.75,24.036){\epsfysize=7.94cm \epsfxsize=8.03cm
\epsffile[161 308 509 655]{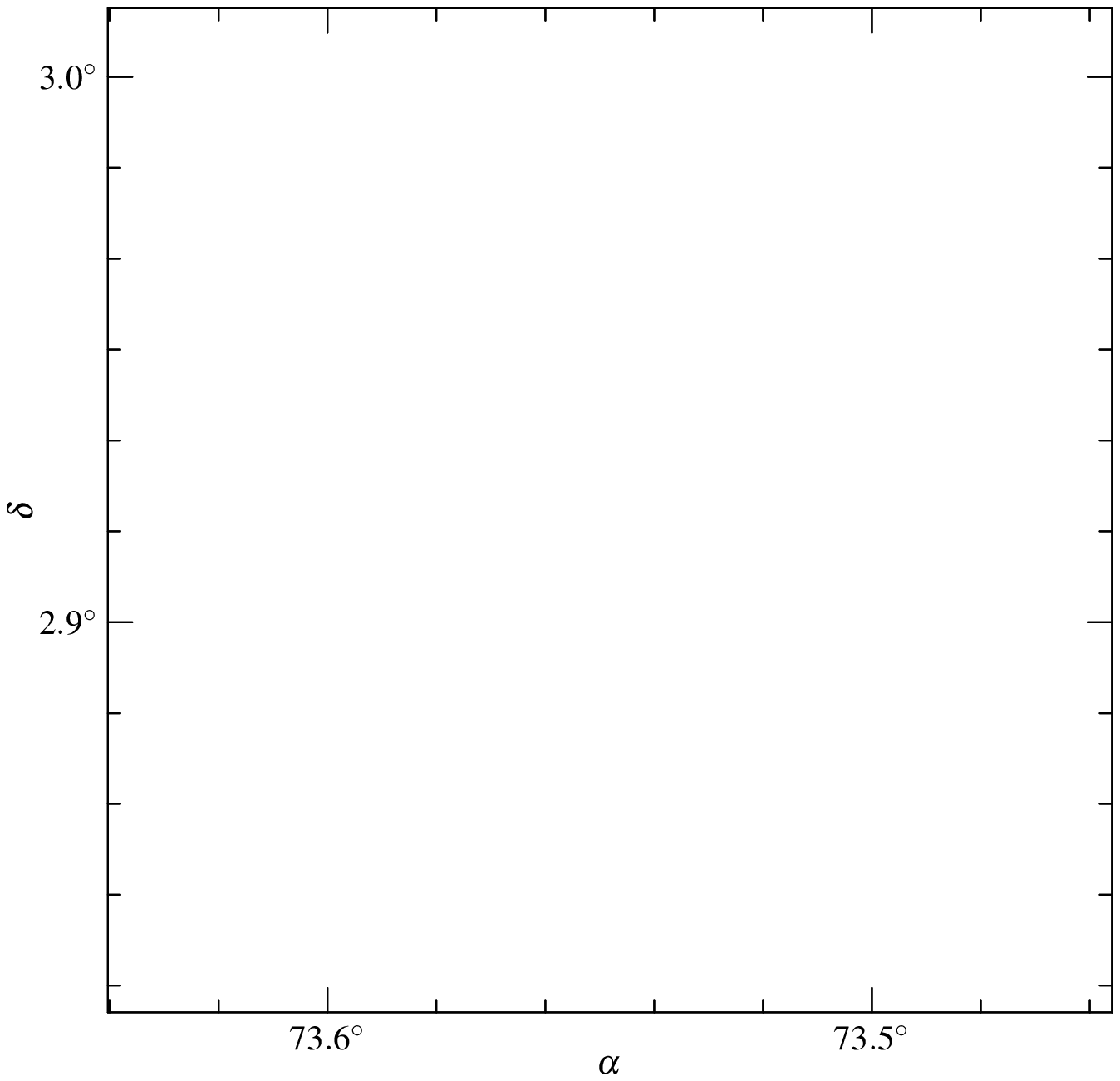}}

\rput[tl]{0}(9.36,24.036){\epsfysize=7.94cm \epsfxsize=8.03cm
\epsffile[161 308 509 655]{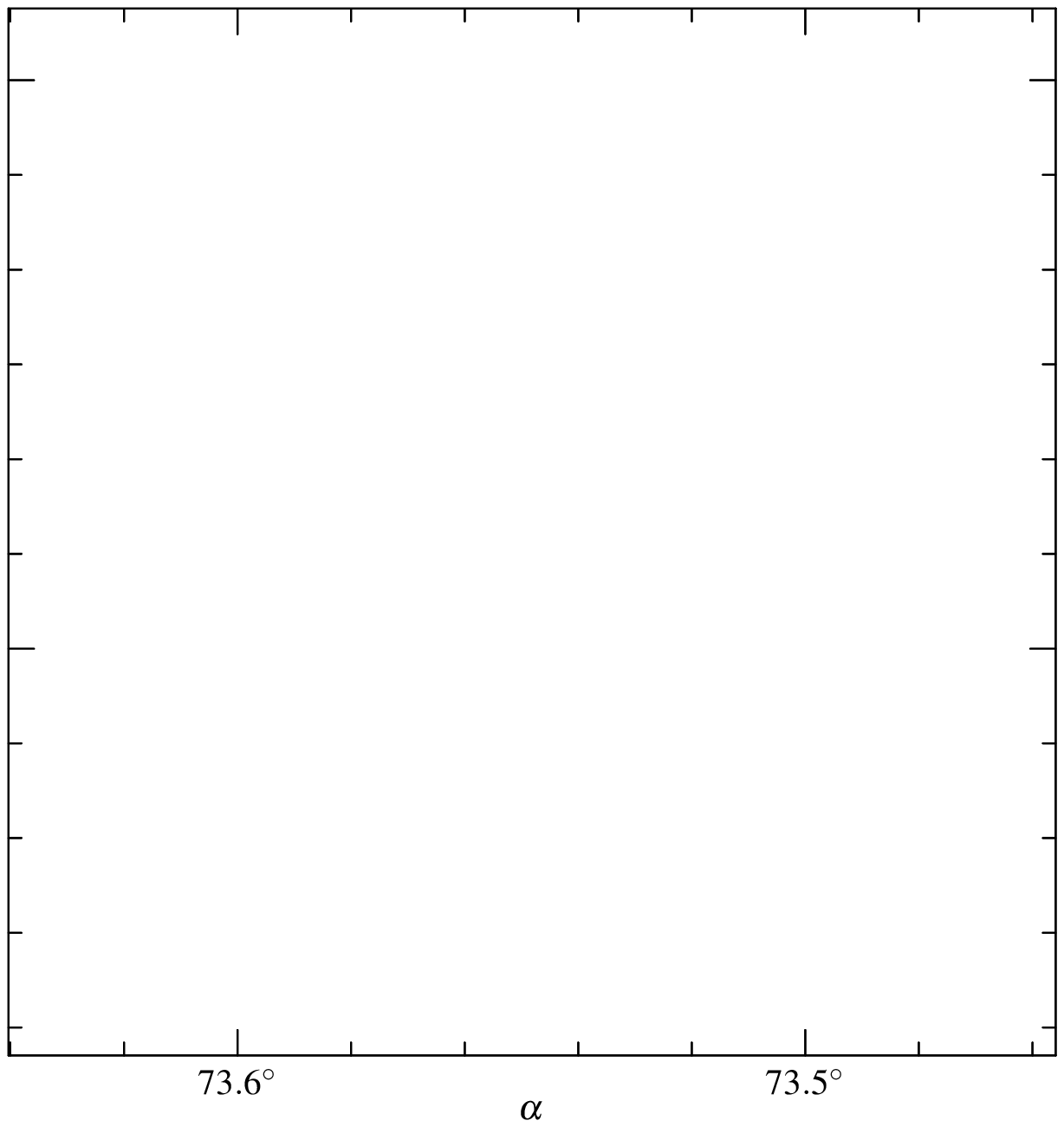}}

% center of image (a): 20.05,4.55 cm = 1064,1064 pix; img size 360 pix;
% center of sector 1059,1076 pix = 4.65,20.31 cm;
% lines between r=3.3-4cm, angles 274 deg, 354 deg
\psline[linestyle=dashed,linewidth=0.5pt]{-}%
(7.93,19.97)(8.63,19.89)
\psline[linestyle=dashed,linewidth=0.5pt]{-}%
(4.88,17.02)(4.93,16.32)

\psline[linewidth=0.5pt]{|-|}(1.30,16.6)(2.38,16.6)
\rput[bc]{0}( 1.84,16.85){300 kpc}

\psline[linewidth=0.5pt]{|-|}(9.90,16.6)(10.98,16.6)
\rput[bc]{0}(10.44,16.85){300 kpc}

\rput[bl]{0}(1.3, 23.5){\large\bi a}
\rput[bl]{0}(9.9,23.5){\large\bi b}

\endpspicture

\caption{({\em a}) ACIS 0.5--2 keV image of A520 smoothed with a
  $\sigma=4''$ Gaussian. Point sources are removed. A bow shock is
  propagating in the SW direction from the cluster center. Dashed lines mark
  a sector used for radial profiles across the shock shown in Fig.\
  \ref{fig:profs}. ({\em b}) \vla\ 1.4 GHz contours from Govoni et al.\
  (2001) overlaid on the X-ray image. Discrete sources not related to
  the diffuse emission are not removed from the
  radio image. FWHM is $15''$; contours are spaced by a factor of 2.}
\label{fig:img}
\end{figure*}
%%%%%%%%%%%%%%%%%%%%%%%%%%%%%%%%%%%%%%%%%%%%%%%%%%%%%%%%%%%%%%%%%%%%%%%%%%

\section{X-RAY ANALYSIS}
\label{sec:data}

A520 was observed with \chandra\ ACIS-I twice.  Results from the first,
short (9 ks) exposure taken in October 2000 were presented in G04. In
December 2003, the cluster was re-observed for 67 ks. We analyze this
dataset here and omit the earlier exposure for simplicity. The data were
prepared and cleaned in a standard manner; the instrument responses for
spectral analysis were generated as described in Vikhlinin et al.\
(2004). The background was modeled using the blank-sky dataset as described
in Markevitch et al.\ (2003b). We encountered a minor complication in the
form of a slowly-varying 10\% background excess over the quiescent level at
$E<8$ kev, somewhat similar to (but smaller than) that described in
Markevitch et al.\ (2002).  In the image area free of cluster emission, and
in the energy interval 0.8--9 keV that we used for spectral fitting, the
spectrum of the excess was well-modeled by a power law with photon index
$-0.7$ (without applying the mirror effective area or CCD efficiency to the
model).  We assumed this component to be spatially uniform and included it
in the fits.  The uncertainty of this component propagates into errors of
best-fit temperatures that are smaller than the statistical errors for the
interesting cluster regions.

Fig.\ \ref{fig:img}{\em a}\/ shows a smoothed ACIS image of the cluster. As
seen from the X-ray as well as weak lensing data (Clowe et al.\ in prep.),
the cluster undergoes a merger along the NE-SW direction.  A bright
irregular structure southwest of center consists of dense, cool pieces of a
cluster core that has been broken up by ram pressure as it flew in from the
NE direction (Markevitch et al.\ in prep). In front of this structure is a
prominent bow-shaped brightness edge perpendicular to the merger
direction. This was suggested by the earlier, shorter, exposure; the new
observation provides sufficient statistics to determine that it is a shock
front.

\begin{figure*}[t]
\pspicture(0,17.8)(18.5,24.0)
%\psgrid(0,15)(18,24)

\rput[tl]{0}(0.2,24){\epsfysize=6.0cm \epsfclipon
\epsffile[19 180 520 680]{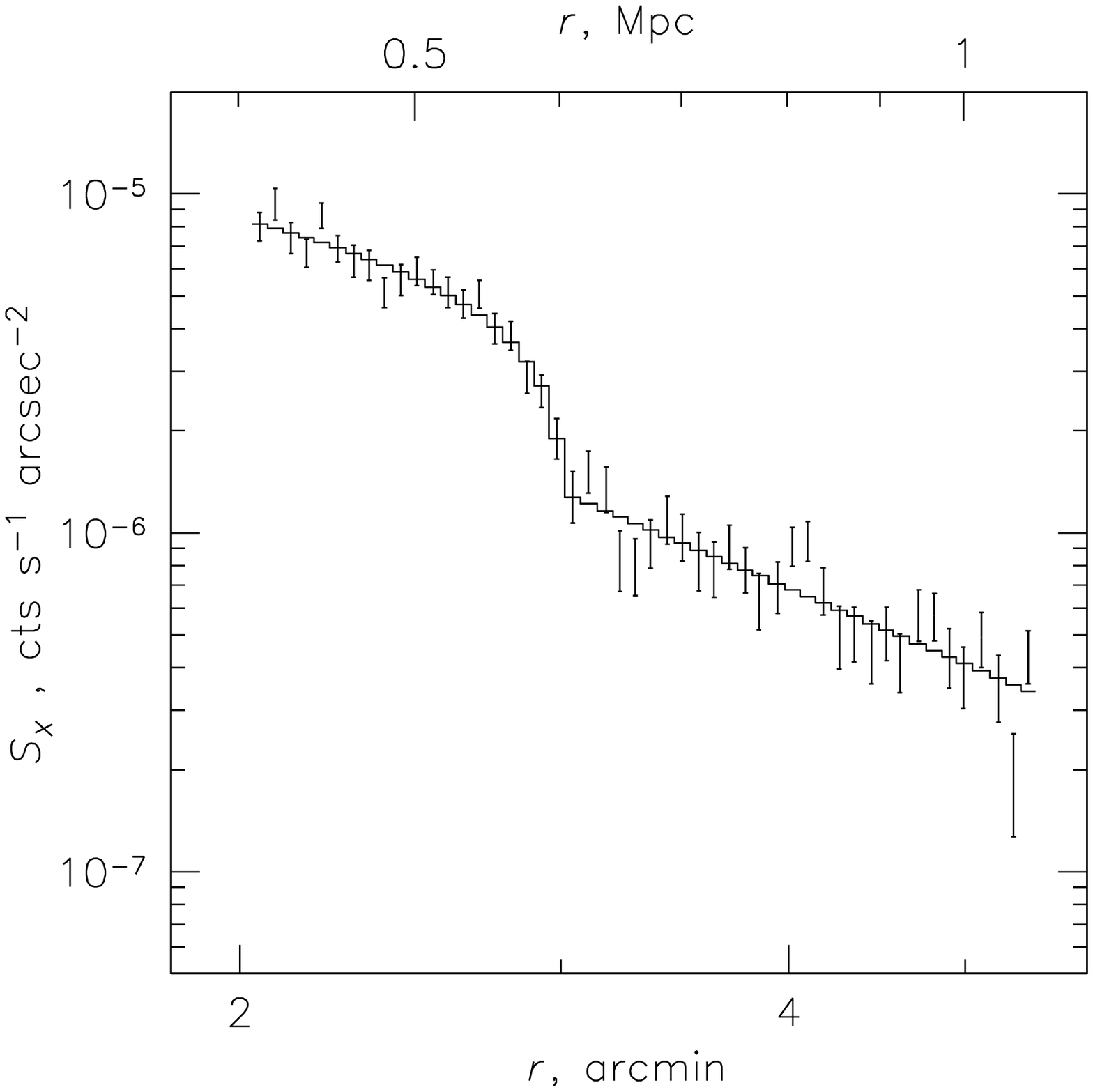}}

\rput[tl]{0}(6.5,24){\epsfysize=6.0cm \epsfclipon
\epsffile[19 180 482 680]{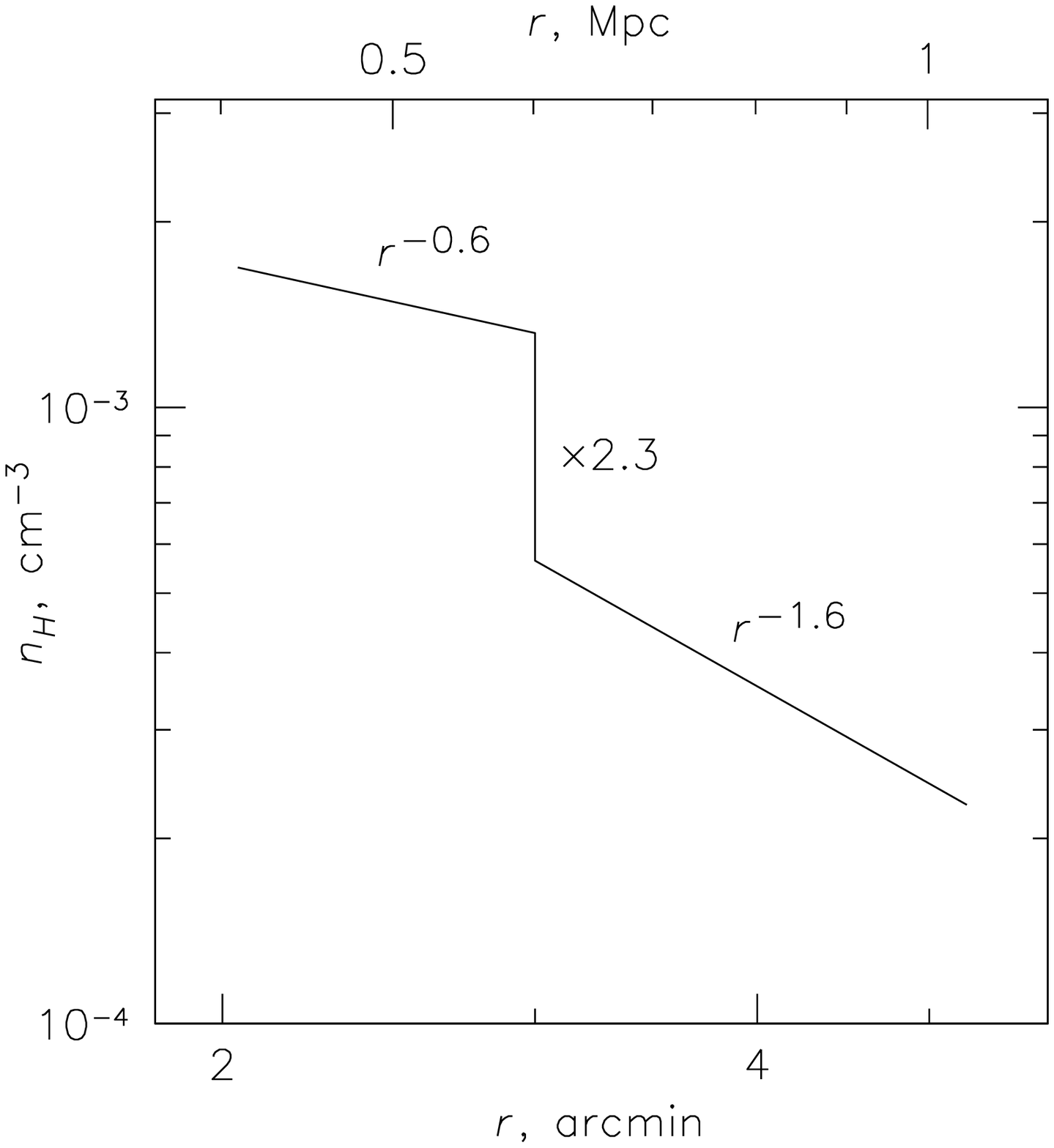}}

\rput[tl]{0}(12.3,24){\epsfysize=6.0cm \epsfclipon
\epsffile[19 180 482 680]{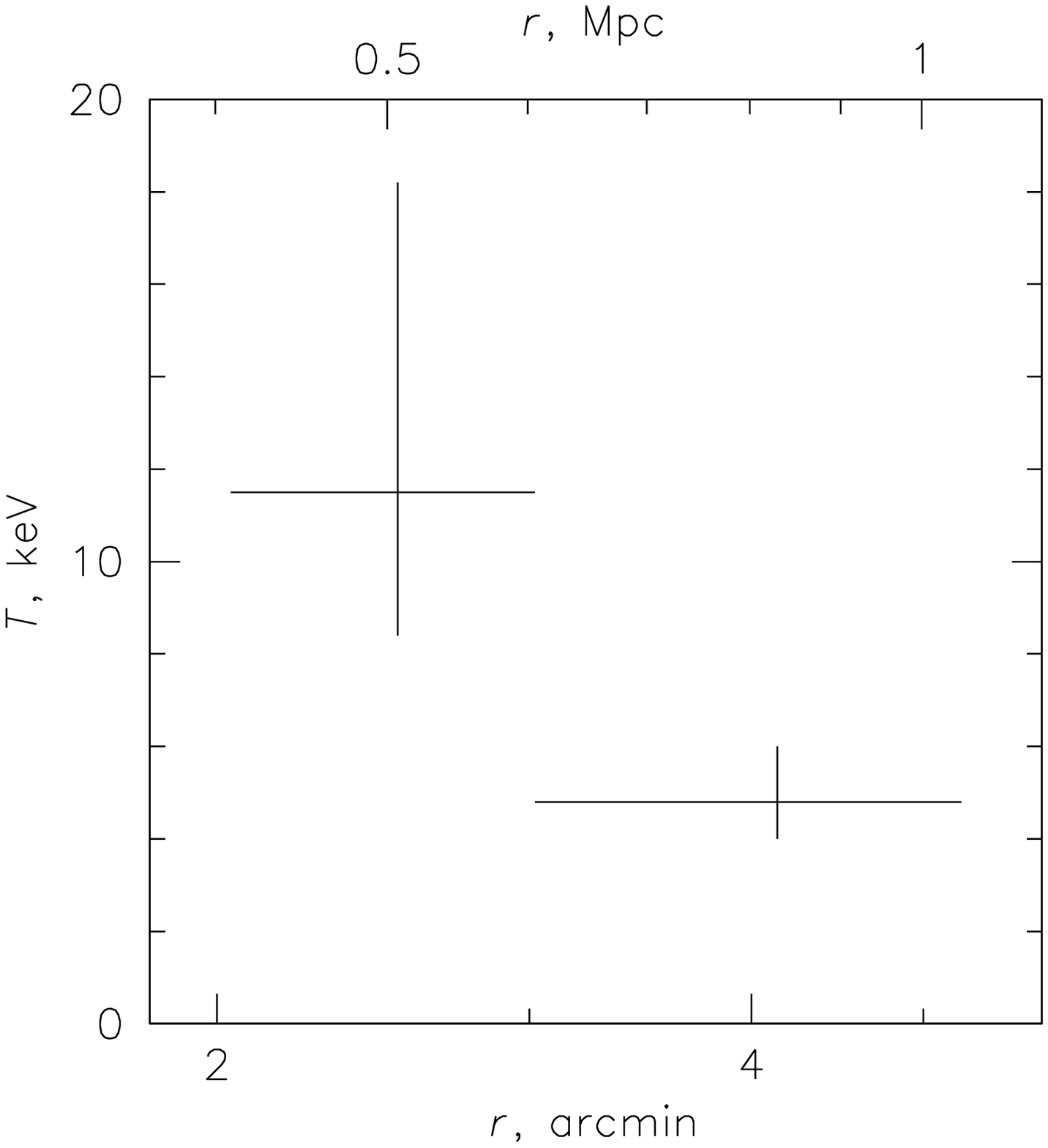}}

\rput[bl]{0}(5.6, 22.9){\large\bi a}
\rput[bl]{0}(11.5,22.9){\large\bi b}
\rput[bl]{0}(17.3,22.9){\large\bi c}

\endpspicture

\caption{({\em a}) The 0.5--2 keV brightness profile across the bow shock in a
  sector marked in Fig.\ \ref{fig:img}{\em a}\/ (see text). Errors here are
  $1\sigma$; histogram shows the best-fit model whose radial density profile
  is given in panel ({\em b}). ({\em c}) Temperatures across the shock
  (errors are 90\%). The $r$\/ axis shows a distance from the shock's center
  of curvature, which is near the centroid of the cluster.}
\label{fig:profs}
\end{figure*}

Fig.\ \ref{fig:profs}{\em a}\/ shows a radial brightness profile across this
edge, extracted in a sector marked in Fig.\ \ref{fig:img}{\em a} and
excluding regions affected by the dense core structures.  The energy band
0.5--2 keV was chosen to minimize the dependence of the X-ray brightness on
temperature. The profile has the characteristic shape of a projected sharp
spherical density discontinuity.  We therefore fit it with such a radial
density profile (two power laws with a jump), shown in panel {\em b}\/ and
in projection as a histogram in panel {\em a}. The best-fit amplitude of the
density jump, after a small (3\%) correction for the measured temperature
difference across the edge, is $x\equiv \rho_2/\rho_1=2.3\pm0.3$.  The gas
temperatures in the same sector are $T_1=4.8^{+1.2}_{-0.8}$ keV and
$T_2=11.5^{+6.7}_{-3.1}$ keV for the low and high density side of the edge,
respectively. The latter temperature value is ``deprojected'', for which we
used the best-fit density model to subtract the (small) contribution of the
outer cooler gas on the line of sight near the shock.

The sign of the temperature jump confirms that the edge is indeed a shock
front. From the Rankine-Hugoniot shock conditions, we can use the above
density and temperature jumps to obtain two independent estimates of its
Mach number. They are $M=2.1^{+0.4}_{-0.3}$ and $2.2^{+0.9}_{-0.5}$,
respectively, in good mutual agreement; we will use the more accurate value
from the density jump. From the gas the temperatures and the Mach number,
the velocities of the the pre-shock and post-shock gas flows in the shock
reference frame are approximately 2300 \kms\ and 1000 \kms, respectively.

%%%%%%%%%%%%%%%%%%%%%%%%%%%%%%%%%%%%%%%%%%%%%%%%%%%%%%%%%%%%%%%%%%%%%%%%%%
%%%%%%%%%%%%%%%%%%%%%%%%%%%%%%%%%%%%%%%%%%%%%%%%%%%%%%%%%%%%%%%%%%%%%%%%%%

\section{DISCUSSION}

\subsection{Radio edge}
\label{sec:halo}

A520 has a prominent radio halo (Govoni et al.\ 2001), whose brightness
contours are overlaid on the X-ray image in Fig.\ \ref{fig:img}{\em b}.  G04
already noted a remarkable coincidence of the SW edge of the radio halo with
the shock-like X-ray feature, although at that time it was not clear whether
the feature is a shock front. A similar extension of the radio halo edge to
the bow shock is seen in \1e\ (Markevitch et al.\ 2002; G04). In another
merging cluster, A665, a ``leading'' edge of the halo also corresponds to a
region of hot gas that is probably behind a bow shock, although the X-ray
image of A665 does not show a gas density edge at the shock, probably due to
an unfavorable viewing angle (Markevitch \& Vikhlinin 2001; G04).  The
overall structure of the halo in A520 may even suggest two distinct
components, a mushroom with a stem and a cap, where the main stem component
goes across the cluster and the cap ends at the bow shock.
The main component appears to be in the region of the cluster where one
expects relatively strong turbulence (G04). In this paper, we concentrate on
the halo edge.

\subsubsection{Shock acceleration}
\label{sec:accel}

One possible explanation for the radio edge is acceleration of electrons to
ultrarelativistic energies by the shock.  The shock should generate
electrons with an energy spectrum $dN/d\gamma=N_0 \gamma^{-p}$
with 
\begin{equation}
p=\frac{x+2}{x-1}
\label{eq:p}
\end{equation}
via first-order Fermi acceleration (e.g., Blandford \&
Eichler 1987); $p\simeq 3.3$ for our shock with $x\simeq 2.3$.  The
synchrotron emission should have a spectrum $I_\nu \propto \nu^{-\alpha}$
with $\alpha=(p-1)/2 \simeq 1.2$ right behind the shock front.  However,
these electrons are short-lived because of IC and synchrotron energy losses,
and their spectrum will quickly steepen.  The respective electron lifetimes,
$t_{\rm\,IC}$ and $t_{\rm syn}$, are
\begin{equation}
t_{\rm\,IC} = 2.3\times 10^{12}\; \gamma^{-1}\, (1+z)^{-4}\; {\rm yr}
\label{eq:tic}
\end{equation}
and
\begin{equation}
t_{\rm syn} = 2.4\times 10^{13}\; \gamma^{-1}\,
      \left(\frac{B}{1\,\mu G}\right)^{-2}\; {\rm yr}
\label{eq:tsyn}
\end{equation}
(e.g., Sarazin 1999). IC losses dominate for $B<3\,\mu G$; other losses are
negligible for our range of energies and fields.  For a power-law electron
spectrum with $p=2-4$ (expected for shocks with $M>1.7$), the contribution
of different $\gamma$ at a given synchrotron frequency has a peak at
\begin{equation}
\gamma_{\rm peak} \approx 10^4\, \left(\frac{\nu}{1\,{\rm GHz}}\right)^{1/2}
   \left(\frac{B}{1\,\mu G}\right)^{-1/2}.
\label{eq:gmax}
\end{equation}
The exact value depends only weakly on $p$\/ in the interesting range, and
the emission at a given frequency comes from a relatively narrow interval of
$\gamma$.  Assuming $B\sim 1\,\mu G$, the lifetime for electrons with
$\gamma\sim 1.2\times 10^4$ that emit at 1.4 GHz in A520 is $\sim 10^8$ yr.
Thus, given the 1000 \kms\ velocity of the downstream flow
(\S\ref{sec:data}) that carries these electrons away from the shock, the
width of the synchrotron-emitting region should only be about 100 kpc,
beyond which the electrons cool out of the 1.4 GHz band. At lower
frequencies, the cooling is slower and this region will be wider (we discuss
this in more detail in \S\ref{sec:b}). This scale is an order of magnitude
smaller than the size of the halo, so the whole halo cannot be produced by
particles accelerated at this shock. This is true for merger shocks in
general (e.g., Brunetti 2003).  However, the cap-like part of the radio halo
appears to have just the right width, $\Delta r\lax 100$ kpc (considering
the finite angular resolution).  Thus with the available data, this region
is not inconsistent with shock acceleration and may be a ``radio gischt'',
using the classification of Kempner et al.\ (2004).  While the relativistic
particles in this structure cool down soon after the shock passage, some may
later be picked up and re-accelerated as they reach the turbulent region
behind the subcluster core, where the stem-like halo component forms.

\subsubsection{Compression of fossil electrons}
\label{sec:compr}

However, the efficiency with which collisionless shocks in clusters can
accelerate particles%
\footnote{The acceleration efficiency is defined as the inverse of the
average time it takes to double the particle's energy.}
is unknown, and may be insufficient to generate the observed radio
brightness. The radio edge in A520 offers an interesting prospect for
constraining it. If the acceleration efficiency is low, the observed radio
edge may alternatively be explained by an increase in the magnetic field
strength and the energy and density of the pre-existing relativistic
electrons simply due to the gas compression at the shock. Such pre-existing
particles may be, for example, secondary electrons generated by relativistic
protons (e.g., Dennison 1980) or cooled primary electrons accumulated from
previous merger events (e.g., Sarazin 1999).  This is similar to the
scenario proposed by En{\ss}lin \& Gopal-Krishna (2001) for radio relics,
which may be regions of fossil radio plasma re-energized by a recent merger
shock passage.  In this model, the pre-existing electrons must produce
diffuse radio emission {\em in front of}\/ the bow shock, whose intensity
and spectrum may be predicted from the shock compression factor and the
post-shock radio spectrum, as given below.

For a rough estimate, we assume that as the gas crosses the shock surface,
its tangled magnetic field is compressed isotropically (indeed, observations
at the Earth's bow shock show that a shock passage strengthens the field
whether it is parallel or perpendicular to the shock, e.g., Wilkinson 2003
and references therein). The average field strength $B$\/ then increases by
a factor 
\begin{equation}
B\propto x^{2/3}
\label{eq:b}
\end{equation}
(and the energy density as $x^{4/3}$), where $x$\/ is the gas density jump.
We further assume for simplicity that relativistic particles do not have a
significant velocity component along the field lines. This shouldn't be too
far from the truth, because in a highly nonuniform field, the particles
spend most of their time in the field bottlenecks and mirrors where the
component of their momentum perpendicular to the field indeed dominates. The
shock passage would then cause the particles to spin up as \mbox{$\gamma
\propto B^{1/2}$} (the adiabatic invariant). For a power-law fossil electron
spectrum of the form
\begin{equation}
\frac{dN}{d\gamma}=N_0\, \gamma^{-\delta},
\label{eq:foss}
\end{equation}
such a shift in energy would preserve the slope of the spectrum and increase
its normalization proportional to $B^{\,(\delta-1)/2}$. In addition, the
number density of relativistic electrons increases by another factor of
$x$\/ due to the compression, so the electron spectrum normalization after
the shock passage will change as (given eq.\ \ref{eq:b})
\begin{equation}
N_0 \propto x^{(\delta+2)/3}.
\label{eq:ncompr}
\end{equation}
(Note that we recover the relativistic adiabat for the particles: across the
shock, their volume energy density changes as $P\propto x \gamma \propto
x^{4/3}$.)  For a power-law electron spectrum, the synchrotron surface
brightness at a given frequency is $I_\nu \propto \int N_0\,
B^{\,(\delta+1)/2} dl$\/ (e.g., Rybicki \& Lightman 1979), where the
integral is along the line of sight.  For our viewing geometry with the
shock surface parallel to the line of sight, it should exhibit a jump at the
shock
\begin{equation}
I_\nu \propto x\,B^{\,\delta} \propto x^{\,\frac{2}{3}\delta+1}. 
\label{eq:icompr}
\end{equation}
Thus, for our shock with $x=2.3$, if the radio edge is due to the
compression only, there must be a pre-shock radio emission with the same
spectrum as post-shock but fainter by a factor of $7-20$ for $\delta=2-4$,
respectively (in practice, $\delta$\/ can be determined from the post-shock
radio spectrum).

The assumption of a power law spectrum can be relaxed somewhat. The
frequency at which the particle emits most of its synchrotron radiation
scales across the density jump as
\begin{equation}
\nu_{\rm peak} \propto B\, \gamma^2 \propto x^{4/3},
\end{equation}
or by a factor of 3 for the A520 shock. Even if the spectrum deviates from a
pure power law, the pre-shock spectrum can still be predicted directly from
the post-shock spectrum at the frequencies scaled down as given above.

Given the simplifying assumptions about the change of the magnetic field and
particle energy at the shock, the above is of course only a rough estimate,
but it should not be too far off. One can obtain estimates within a factor
of 2 for the $I_\nu$ increase under very simple alternative assumptions
about the field structure.

\subsubsection{Re-acceleration of fossil electrons}
\label{sec:reaccel}

In addition to the two mechanisms discussed above, there is a third
possibility, namely, re-acceleration of the fossil electrons by the shock.
In the linear acceleration regime, the post-shock electron spectrum will be
(e.g., Blandford \& Eichler 1987; Micono et al.\ 1999):
\begin{equation}
\frac{dN}{d\gamma}=(p+2)\,\gamma^{-p}
  \int_{\gamma_{\rm min}}^\gamma
  \frac{dN_{\rm f}}{d\gamma}\, \gamma^{\,p-1}\, d\gamma,
\end{equation}
where $p$\/ is from eq.\ (\ref{eq:p}), $\gamma_{\rm min}$ is the minimum
initial energy of the particles that can be accelerated at the shock (or
present in the fossil population, whichever is higher), and $dN_{\rm
f}/d\gamma$ is the fossil electron spectrum. If the latter is a power law
given by eq.\ (\ref{eq:foss}), and is steeper than the one produced by
direct shock acceleration (i.e., $\delta>p$), the post-shock spectrum will
acquire the slope $p$\/ and its normalization will increase by a large
factor dependent on $\gamma_{\rm min}$. In practice, this case will probably
be indistinguishable from the direct shock acceleration considered above.
If the fossil spectrum is flatter with $\delta<p$, the post-shock spectrum
will preserve the slope $\delta$\/ while its normalization will increase as
\begin{equation}
N_0 \propto \frac{3x}{x+2-\delta\,(x-1)}.
\end{equation}
A comparison of this factor to the one derived for the compression (eq.\
\ref{eq:ncompr}) confirms the intuitive expectation that re-acceleration
would provide a bigger boost to the fossil electron spectrum than simple
compression. For example, for $\delta=2-3$ (a flatter slope than $p\approx
3.3$ for our shock), an increase due to the compression is by factors of
$\approx 3-4$ for our shock, while the re-acceleration increase is by
factors of $\approx 4-17$, respectively. We should note that a boost of the
fossil electron spectrum due to the magnetic field compression should occur
regardless of the presence of shock acceleration, but whether or not these
two effects are additive probably depends on the exact processes on the
microscopic level.

\subsubsection{Observational prospects}
\label{sec:obs}

The above estimates show that the least efficient way to create the radio
edge is the increase of the energy of the pre-existing relativistic
electrons due to the compression of the magnetic field by the shock. In such
a scenario, those fossil electrons should also generate synchrotron emission
in front of the shock.  If future sensitive measurements do not detect a
pre-shock radio emission at the level predicted for simple compression, it
would mean that the shock generates relativistic electrons and/or a magnetic
field, as opposed to simply compressing them. Observations of interplanetary
shocks of a similar strength and high thermal-to-magnetic pressure ratio
relevant for cluster plasma seem to be consistent with a simple compression
of the field (Russell \& Greenstadt 1979), so magnetic field generation is
unlikely.  Thus such a non-detection would provide a direct estimate of the
shock's particle acceleration efficiency.

While the expected pre-shock emission may be below the current measurement
sensitivity, one can get an initial idea of the origin of the electrons in
the cap of the halo by measuring the radio spectrum immediately behind the
edge. If the electrons are shock-accelerated directly, the radio spectral
slope should be $\alpha=1.2$. Within the range of possibilities that we have
considered, a steeper spectrum would point to compressed fossil electrons,
and a flatter spectrum would indicate fossil electrons with a flat initial
energy spectrum, either compressed or re-accelerated.

Because the bow shock is spatially separated from the turbulent area further
downstream (except for the region around the small dense core fragments) and
there is no reason to expect significant turbulence and additional
acceleration in that intermediate region, the cap-like structure is likely
to exhibit a measurable spectral difference from the main halo. Within the
100 kpc-wide strip along the shock, the spectrum should quickly steepen
starting from $\alpha=1.2$ (or other, probably not very different, value as
discussed above).  If the region is unresolved, the resulting mixture would
have a volume-averaged slope $\bar{\alpha}\approx \alpha+1/2$ (Ginzburg \&
Syrovatskii 1964) which is significantly steeper than $\alpha\simeq 1-1.2$
observed on average in most halos (e.g., Feretti 2004), the bulk of which is
probably continuously powered by turbulence.  Interestingly, Feretti et al.\
(2004) found that the presumed post-shock region in A665 indeed exhibits the
steepest radio spectrum in the spectral index map of the cluster, which is
consistent with the above two-component cap + stem picture.

%%%%%%%%%%%%%%%%%%%%%%%%%%%%%%%%%%%%%%%%%%%%%%%%%%%%%%%%%%%%%%%%%%%%%%%%%%
\begin{figure}[t]
\pspicture(0,16.2)(8.8,23.8)
%\psgrid(0,15)(8.5,24)

\rput[tl]{0}(0.2,24){\epsfysize=8cm
\epsffile{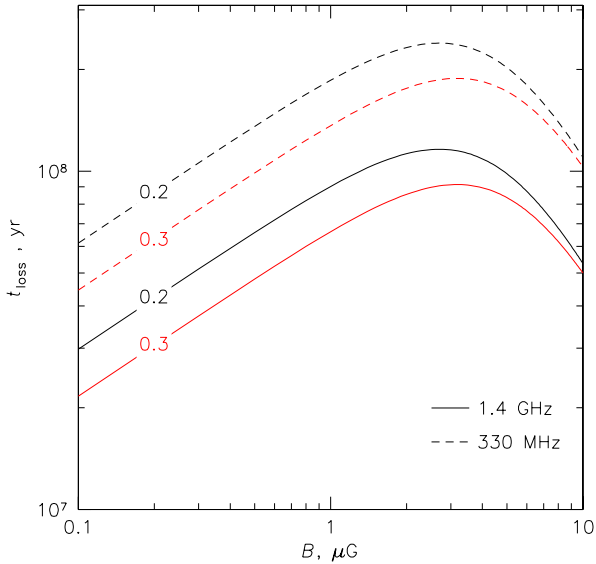}}

\endpspicture

\caption{Lifetime of the relativistic electrons that contribute the most
  to the synchrotron emission at a certain frequency, for a $p=3-4$ electron
  spectrum, as a function of $B$. Two frequencies are shown as solid and
  dashed lines.  Black and red lines correspond to two different redshifts
  ($z=0.2$ for A520 and $z=0.3$ for \1e).}
\label{fig:b}
\end{figure}
%%%%%%%%%%%%%%%%%%%%%%%%%%%%%%%%%%%%%%%%%%%%%%%%%%%%%%%%%%%%%%%%%%%%%%%%%%

\subsection{Magnetic field estimate}
\label{sec:b}

Regardless of the exact origin of the relativistic electrons responsible for
the radio edge, the distinct ``gischt'' in A520, and perhaps a similar
region in \1e, provide an interesting possibility for estimating the
magnetic field strength behind the shock. As discussed above, IC and
synchrotron cooling cause the electron spectrum to steepen and the electrons
to drop out of the radio image on timescales of order $10^8$ yr. This
timescale depends on $B$\/ as shown in Fig.\ \ref{fig:b}, which combines
eqs.\ (\ref{eq:tic}, \ref{eq:tsyn}, \ref{eq:gmax}). It gives the lifetime of
the electrons that contribute the most at a given frequency, for a $p=3-4$
spectrum and an interesting range of $B$.  We assume that the bow shock
causes a momentary increase in electron energy and $B$. The post-shock gas
flows away from the shock with a known velocity and effectively ``unrolls''
the time evolution of the electron spectrum along the spatial coordinate
(assuming no diffusion, see \S\ref{sec:diff}, and neglecting for this simple
estimate the change of the post-shock velocity with coordinate). Thus the
width of the ``gischt'', if it is resolved, can give us the measure of the
magnetic field. In practice, such a measurement needs to be done at more
than one frequency in order to determine the spectrum of the electrons for a
more precise calculation of $\gamma_{\rm peak}$ (and of course to verify
that the electrons cool as predicted at different frequencies, that is, no
acceleration occurs after the shock passage). In addition, the radio angular
resolution has to be sufficient to perform at least a rough deprojection of
the spherical edge of the bow shock (cf.\ Figs.\ \ref{fig:profs}{\em a,b})
and exclude the (possibly turbulent) region around the dense core remnant
(Fig.\ \ref{fig:img}{\em a}). A cooling time $2\times 10^7$ yr (the minimum
interesting time) corresponds to the 6\as\ distance from the front, which
gives the minimum resolution requirement. The available single-frequency
radio data (Fig.\ \ref{fig:img}{\em b}) do not have the needed signal to
noise ratio or resolution, but are not inconsistent with $B$\/ of order a
$\mu G$.  Such a measurement may be easier in \1e\ than in A520, as the
shock there is more prominent and farther away from the turbulent area.

This method is reminiscent of that of Harris \& Romanishin (1974) which
compares the hard X-ray IC and radio synchrotron luminosities, in that it
also combines the IC and synchrotron emission. The field values derived from
the IC/synchrotron ratio have persistently been lower than those derived
from Faraday rotation (see Carilli \& Taylor 2002 for a review). One of the
proposed explanations is that the hard X-ray and radio emissions come from
different electron populations in regions with different fields (e.g.,
Rudnick 2000; Petrosian 2001; Brunetti et al.\ 2001). Our method would be
free from this complication, because we essentially estimate the IC losses
using the time evolution of the synchrotron emission from exactly the same
electrons.  Our method can distinguish among the values in the controversial
interval of $B\sim 0.1-3\,\mu G$, although, as seen from Fig.\ \ref{fig:b},
it cannot give a unique value of $B$, because $t_{\rm loss}$ is not a
monotonous function of $B$. Note also that the magnetic field behind the
shock would be amplified as discussed in \S\ref{sec:compr}.

\subsection{Diffusion of relativistic particles}
\label{sec:diff}

The above argument assumes that relativistic particles do not diffuse
through the gas away from their place of origin. Diffusion with a velocity
less than the 1000 \kms\ shock velocity would not alter the picture
qualitatively.  The diffusion speed is not expected to exceed the Alfv\'en
velocity of $\approx 50\; (B/1\,\mu G)$ \kms\ (e.g., Jaffe 1977), because
faster-diffusing particles should be slowed down by self-generated Alfv\'en
waves.
If in reality this mechanism is not effective and the diffusion is much
faster, a factor of $M=2.1$ or more above the sound speed, it will spread
the radio edge into the pre-shock region. Therefore, the study of the radio
edge also provides an interesting opportunity to detect or place a direct
limit on the diffusion of relativistic particles.  

Diffusion would create pre-shock radio emission which may be confused with
that from any fossil particle population discussed in \S\ref{sec:compr}.
However, unlike the fossil particles which do not know anything about this
shock, diffusion would create a narrow strip clearly related to the shock
(for any reasonable diffusion rates). Thus, if any pre-shock emission is
ever detected, it should not be difficult to separate these two
possibilities.

\section{SUMMARY}

The \chandra\ observation of A520 reveals a prominent bow shock with
$M=2.1$, which is only the second clear example of a substantially
supersonic merger shock front besides \1e. The shock coincides with an
apparent leading edge of the radio halo. In light of this coincidence, we
discuss possible explanations for the radio edge, which include direct
acceleration of relativistic electrons by the shock, or energizing
pre-existing electrons via shock re-acceleration or the compression of the
magnetic field.  Both models make testable predictions. In particular, the
compression model predicts the existence of pre-shock radio emission at a
level about 1/10--1/20 of the post-shock brightness. If one determines which
of the models is valid, it could provide a measure of the particle
acceleration efficiency in the cluster merger shocks.  The slope of the
radio spectrum immediately inside the edge may also give an idea of the
nature of the edge, and possibly even test the applicability of the Fermi
acceleration mechanism.

Regardless of the nature of the radio edge, it offers an interesting tool to
measure the magnetic field strength in the post-shock gas (albeit not
unambiguously) and exclude high rates of diffusion of the relativistic
electrons, provided high-resolution radio data. All these measurements rely
on the knowledge of the shock velocity and compression factor obtained from
the X-ray.  The shock fronts in A520 and \1e\ can thus provide unique
information on the microphysics of the intracluster gas.

\acknowledgements

We thank Pasquale Blasi, Tracy Clarke, Alexey Vikhlinin and Dan Harris for
stimulating discussions, and the referee for useful comments.  Support for
this work was provided by NASA contract NAS8-39073, \chandra\ grants
GO2-3164X and GO3-4172X, and by INAF grant D4/03/15.


\begin{references}
\raggedright

\reference{} Blandford, R.~\& Eichler, D.\ 1987, \physrep, 154, 1

\reference{} Brunetti, G., Setti, G., Feretti, L., \& Giovannini, G.\ 2001,
\mnras, 320, 365
% reaccel.

\reference{} Brunetti G., 2003, Proc.\ ``Matter and Energy in Clusters of
Galaxies'', Taiwan, ASP Conf.\ Series, eds.\ S. Bowyer \& C.-Y. Hwang, 301,
349 (astro-ph/0208074)

\reference{} Buote, D. A., 2001, ApJ, 553, L15

\reference{} Carilli, C.~L.~\& Taylor, G.~B.\ 2002, \araa, 40, 319 

\reference{} Dennison, B.\ 1980, ApJ, 239, L93 

\reference{} En{\ss}lin, T.~A.~\& Gopal-Krishna 2001, \aap, 366, 26 

\reference{} Feretti, L. 2002, IAU Symposium, 199, 133 (astro-ph/0006379)

\reference{} Feretti, L. 2004, astro-ph/0406090
% indices for several clusters

\reference{} Feretti, L., Orr{\`u}, E., Brunetti, G., Giovannini, G.,
Kassim, N., \& Setti, G.\ 2004, \aap, 423, 111
% index maps

\reference{} Fujita, Y., Takizawa, M., \& Sarazin, C.~L., 2003, \apj, 584, 190 

\reference{} Fusco-Femiano, R., Orlandini, M., Brunetti, G., Feretti, L., 
Giovannini, G., Grandi, P., \& Setti, G.\ 2004, \apjl, 602, L73 
% coma IC field

\reference{} Ginzburg, V.~L.~\& Syrovatskii, S.~I.\ 1964, The Origin of 
Cosmic Rays (New York: Macmillan)

\reference{} Govoni, F., Feretti, L., Giovannini, G., B\"ohringer, H.,
Reiprich, T.H., \& Murgia, M., 2001, A\&A, 376, 803
% a520 halo

\reference{} Govoni, F., Markevitch, M., Vikhlinin, A., VanSpeybroeck, L.,
Feretti, L., \& Giovannini, G. 2004, ApJ, 605, 695 (G04)

\reference{} Harris, D.~E.~\& Romanishin, W.\ 1974, \apj, 188, 209 

\reference{} Harris, D. E., Kapahi, V. K., \& Ekers, R. D. 1980, A\&AS, 39,
215 

\reference{} Jaffe, W.~J.\ 1977, \apj, 212, 1 
 
\reference{} Kempner, J.~C., Blanton, E.~L., Clarke, T.~E., En{\ss}lin,
T.~A., Johnston-Hollitt, M., \& Rudnick, L.\ 2004, in The Riddle of
Cooling Flows, eds.\ T. Reiprich, J. Kempner, \& N. Soker (astro-ph/0310263)

\reference{} Liang, H., Hunstead, R. W., Birkinshaw, M., \& Andreani, P.
2000, ApJ, 544, 686

\reference{} Markevitch, M., Gonzalez, A.~H., David, L., Vikhlinin, A.,
Murray, S., Forman, W., Jones, C., \& Tucker, W.\ 2002, \apjl, 567, L27

\reference{} Markevitch, M., Sarazin, C.~L., \& Vikhlinin, A.\ 1999, \apj,
521, 526

\reference{} Markevitch, M., \& Vikhlinin, A., 2001, ApJ 563, 95

\reference{} Markevitch, M., et al., 2003, ApJ, 583, 70 
% cxb

\reference{} Micono, M., Zurlo, N., Massaglia, S., Ferrari, A., \& Melrose,
D.~B.\ 1999, \aap, 349, 323

\reference{} Petrosian, V.\ 2001, \apj, 557, 560

\reference{} Rephaeli, Y.~\& Gruber, D.\ 2002, \apj, 579, 587

\reference{} Rudnick, L.\ 2000, in Cluster Mergers and their Connection to 
Radio Sources, 24th meeting of the IAU, JD10, E22

\reference{} Russell, C.~T.~\& Greenstadt, E.~W.\ 1979, Space Science 
Reviews, 23, 3 
% interplanet. shock, field compression

\reference{} Rybicki, G. B., \& Lightman, A. P. 1979, Radiative processes in
astrophysics (New York: Wiley)

\reference{} Sarazin, C.~L.\ 1999, \apj, 520, 529 

\reference{} Schlickeiser, R., Sievers, A., \& Thiemann, H., 1987, \aap,
182, 21

\reference{} Tribble, P. 1993, MNRAS, 263, 31

\reference{} Vikhlinin, A., Markevitch, M., Murray, S., Jones, C., Forman,
W., \& VanSpeybroeck, L. 2004, ApJ, submitted (astro-ph/0412306)

\reference{} Wilkinson, W.~P.\ 2003, \planss, 51, 629 
% parallel shock near earth 

\end{references}
\end{document}